\documentclass[11pt,twoside]{article}

\usepackage{asp2006}
\usepackage{epsf}
\usepackage{psfig}
\usepackage{lscape}
\usepackage{natbib}

\begin{document}

\title{The CoRoT exoplanet programme: exploring the gas-giant/terrestrial
  planet transition} 

\author{S.\ Aigrain,\altaffilmark{1,2} P.\ Barge,\altaffilmark{3} M.\ Deleuil,\altaffilmark{3} F.\ Fressin,\altaffilmark{4} C.\ Moutou,\altaffilmark{3} D.\ Queloz,\altaffilmark{5} M. Auvergne,\altaffilmark{6} A.\ Baglin\altaffilmark{6} and the CoRoT Exoplanet Science Team}

\altaffiltext{1}{Institute of Astronomy, University of Cambridge, Madingley Road, Cambridge CB3
  0HA, United Kingdom}

\altaffiltext{2}{School of Physics, University of Exeter, Stocker Road, Exeter EX4 4QL,
  United Kingdom}

\altaffiltext{3}{Laboratoire d'Astrophysique de Marseille, Travers\'ee
  du Siphon, Les trois Lucs, BP 8, 13376 Marseille Cedex 12, France}

\altaffiltext{4}{Observatoire de la C\^ote d'Azur, BP 49, F-06304
  Nice, France}

\altaffiltext{5}{Observatoire Astronomique de l'Universit\'e de
  Gen\`eve, 51 chemin des Maillettes, 1290 Sauverny, Switzerland}

\altaffiltext{6}{LESIA, Observatoire de Paris-Meudon, 5 place Jules
  Janssen, 92195 Meudon Cedex, France}

\begin{abstract}
  CoRoT, which was launched successfully on the 27$^{\rm th}$ of
  December 2006, is the first space mission to have the search for
  planetary transits at the heart of its science programme. It is
  expected to be able to detect transits of planets with radii down to
  approximately two Earth radii and periods up to approximately a
  month. Thus, CoRoT will explore the hereto uncharted area of
  parameter space which spans the transition between the gaseous giant
  planets discovered in large numbers from the ground, and terrestrial
  planets more akin to our own. This papers briefly sketches out the
  main technical characteristics of the mission before summarising
  estimates of its detection potential and presenting the data
  analysis and follow-up strategy.
\end{abstract}

\section{The CoRoT mission}

\subsection{Brief history}

CoRoT (Convection, Rotation and Transits) is a modest scale satellite
funded primarily by the French space agency CNES, with additional
contributions from the European Space Agency (ESA), Belgium, Austria,
Germany, Spain and Brazil. The project was first proposed to CNES as
an asteroseismology mission in 1993. An exoplanet programme was added
to the mission soon after the discovery of the first exoplanet around
a Sun-like star \citep{aig_mq95}, and it soon became clear that a long
baseline, high-precision stellar photometry mission had major
potential in exo-planetology as well as in stellar structure and
evolution (see \citealt{aig_blo+06}). The new mission concept was formally
selected by CNES in 2001, and CoRoT was launched on the $27^{\rm th}$
of December 2006 after a successful development and integration phase.

\subsection{The satellite and payload}

The main technical characteristics of the mission, and in particular
of the instrument, are described in detail in the CoRoT instrument
handbook\footnote{See {\tt
    http://corotsol.obspm.fr/web-instrum/payload.param/}.} and the
reference ESA Publication on the pre-launc status of the CoRoT mission
(see Section~\ref{aig_concl}) and are thus only briefly sketched out
here.

\begin{figure}[t]
  \centering
  \plotone{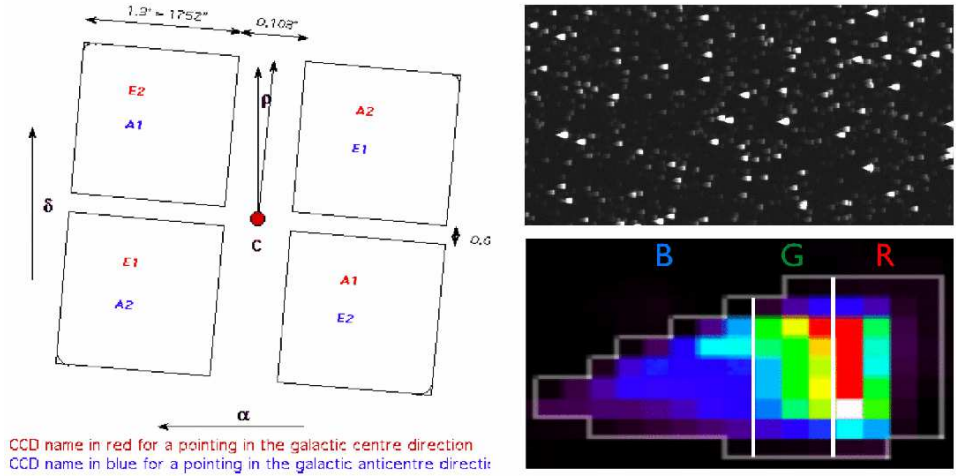}
  \caption{Left: Schematic layout of the focal plane. The CCD names E1 and E2 correspond to the exoplanet field and A1 and A2 to the asteroseismology field. The angles $\alpha$, $\delta$ and $\rho$ are respectively the right ascension and declination of the centre C of the focal plane, and the inclination of the longer side of the asteroseismology or exoplanet field to the meridian. Top right: $\sim 0.6 \times 0.3^{\circ}$ section of a simulated image of the exoplanet CCD. Bottom right: Photometry in the exoplanet channel is conducted by placing a tear-shaped mask over the image of each target (colour coded here according to flux for a star with $T_{\rm eff}=5500$\,K) and integrating the flux within the mask. For targets brighter than $V=15$, the mask is divided into blue, green and red colour channels, containing $\sim 30$, 30 \& 40\% of the flux respectively.
    \label{aig_focal}}
\end{figure}

The payload consists of a 27\,cm aperture afocal telescope equipped
with an extremely high performance baffle to minimise straylight on
the detector. The telescope feeds a focal plane unit of four $2048
\times 2048$ pixel CCDs arranged in a square pattern, with a pixel
scale of $13.5\,\mu$m, corresponding to $2.32$\,arcsec on the sky. Two
of the CCDs are dedicated to asteroseismology, and two to planet
finding, giving rectangular fields of $1.3^{\circ} \times 2.6^{\circ}$
for each program (see Figure~\ref{aig_focal}, left).

The asteroseismology CCDs are set slightly forwards of the focal plane
to create a defocussed image, and are read in frame transfer mode
every second. Up to 5 stars with apparent magnitudes in the range $5.7
< V < 9.5$ are monitored on each CCD, and their light curves are
transmitted to the ground with 32\,s sampling (1\,s on request). In
addition, up to six $30 \times 30$ pixel windows can be transmitted to
the ground every 32\,s. The exoplanet CCDs are read every 32\,s, and
up to 6000 stars per CCD in the range $11.5 < V < 16$ are
monitored. Because of telemetry limitations, aperture photometry is
performed on-board and light curves with 512\,s sampling are
downloaded (32\,s sampling is available for up to 500 stars per
CCD). An objective prism inserted in the light path of the exoplanet
CCDs gives rise to a very low-resolution spectrum at the location of each
star (see Figure~\ref{aig_focal}, right), allowing the generation
of 3-colour light curves for up to 5000 stars with $V <15$ using
appropriately positioned apertures. Small windows spread over the CCD
are used in both fields to monitor the background level. The pointing
of the satellite is adjusted based on astrometry from the
asteroseismology channel.
 
CoRoT uses the multi-mission PROTEUS platform developed by CNES, which
allowed for significant cost savings in the development phase. It was
launched on the $27^{\rm th}$ of December 2006 by a Soyuz II 1-b
launcher from Ba{\" i}konour in Kazakhstan, into a polar orbit with an
altitude of 896\,km and a period of 6714\,s.

\subsection{Observing strategy}
\label{aig_strategy}

To keep scattered light to a minimum implies a Sun angle constraint
which results in two 6-month continuous viewing zones of roughly
$10^{\circ}$\ diameter on opposite sides of the sky along the plane of
the ecliptic. To maximise the number of targets available for planet
finding, two `CoRoT eyes' were selected at the intersections of the
ecliptic and Galactic planes, one towards the Galactic centre (${\rm
  RA}=6^{\rm h}50^{\rm d}$, ${\rm Dec}=0^{\circ}$) and one towards the
Galactic anti-centre (${\rm RA}=18^{\rm h}50^{\rm d}$, ${\rm
  Dec}=0^{\circ}$). CoRoT is restricted to point within these two eyes
only. 

After a commissioning phase lasting roughly one month, the first
science observations will be in the Galactic centre direction, with an
initial run lasting one month to six weeks. After that, CoRoT will
start its regular observing pattern of observing one target field for
5 months continuously (long run), then switching to another field in
the same eye for approximately 21 days (short runs), and then turning
to face the opposite direction. This cycle will be repeated a minimum
of 5 times during the mission lifetime, starting with the Galactic
anti-centre direction in the spring of 2007. 

The choice of each of the target fields for the long runs was a
compromise between the needs of the asteroseismology and the exoplanet
science programmes. Using GAUDI\footnote{See {\tt
    http://sdc.laeff.esa.es/gaudi/}.} \citep{aig_scg+05}, a database
of all stars brighter than $V=9.5$ in the CoRoT eyes, promising
targets for asteroseismology were identified as possible `primary
targets'. The potential planet finding fields around each possible
primary target (allowing for all rotations and translations of the
detectors while keeping the primary target on the asteroseismology
CCDs) were surveyed from the ground using multi-band optical
photometry \citep{aig_dmd+06}, combined with 2MASS, to evaluate the amount
of reddening and the relative fraction of dwarf and giant or early
type stars (the latter, which make an estimated 20\% and 50\% of the
stellar population in the Galactic anti-centre and centre fields
respectively, are not useful targets for transit finding). Also taken
into account when selecting target fields were potential secondary
targets for the asteroseismology fields and potential targets for
additional programmes (see Section~\ref{aig_ap}) in both fields. The
fields of the initial run and the first two long runs are listed in
Table~\ref{aig_fields}.

\begin{table}[t]

  \caption{Fields of the initial run (IR) and the first two long runs (LRc1 and
LRa1 for the Galactic centre and anti-centre direction
respectively). The co-ordinates correspond to the centre of the focal
plane, and the roll angle is the angle between the longer side of the
asteroseismology or exoplanet field and the
meridian (see the focal plane layout in Figure~\ref{aig_focal}).
There may be small changes to the pointings for LRc1 and LRa1 ($<10$'
in $\alpha$, $\delta$ or $\rho$). \label{aig_fields}}

  \smallskip
  \begin{center}
    {\small
    \begin{tabular}{lccc}
      \tableline
      \noalign{\smallskip}
      Name & RA & Dec & Roll angle \\
      & [h~m~s] & [$^{\circ}$~'~''] & [$^{\circ}$] \\
      \noalign{\smallskip}
      \tableline
      \noalign{\smallskip}
      IR & $06~50~25.0$ & $-01~42~00$ & $+09.60$ \\
      LRc1 & $19~23~28.8$ & $+00~28~48$ & $+24.24$ \\
      LRa1 & $06~46~48.0$ & $-00~11~24$ & $+01.92$\\
     \noalign{\smallskip}
     \tableline
    \end{tabular}
  }
  \end{center}
\end{table}

The choice of the target fields for the short runs is primarily driven
by the requirements of the asteroseismology program (to ensure coverage of
all the HR diagram), though a small number of short runs will be
driven by the needs of the additional program.

\section{Transit detection}

The expected scientific performance of CoRoT from the exoplanet point
of view is reported in \citet{aig_bml+06}. This section summarises the
methods that are foreseen to be used to produce light curves and
analyse them, and the results of simulations of the CoRoT fields which
aim to give an estimate of the planet yield of the mission.

\subsection{Light curve generation}

The point spread function (PSF) in the exoplanet field is tear-shaped,
owing to the presence of the prism in the optical path. It is strongly
peaked at the red end and extended in the x- (East-West)
direction. The exact PSF depends on the star's temperature and
position on the CCD, and the on-board software selects the most
appropriate mask for each target from a set of 256 templates, based on
an initial long-exposure image taken at the beginning of each
run. Aperture photometry is then carried out by summing the flux in
the pixels within the mask (see Figure~\ref{aig_focal}, bottom
right). To generate three-colour photometry, the mask is divided into
three sub-masks and the columns falling into each of the sub-masks are
summed separately. Binned averages from small background windows
spread over each CCD are used to monitor the background level. Except
for the 500 oversampled targets per CCD, the aperture photometry from
16 32\,s exposures is summed on-board and individual light curve
points are downlinked to Earth every 512\,s. The raw data received
from the satellite are labelled N0 data. 

The data downlinked from the satellite (labelled N0 data) is further
processed on the ground, first by a generic pipeline which corrects for
anticipated astrophysical and instrumental noise sources and
concatenates the photometry of each object into light curves (N1
data), and then by a pipeline specific to the exoplanet channel which
is designed to correct for unanticipated noise sources identified from
an ensemble analysis of the N1 light curves (N2 data). The N2 data are
the science grade data on which all scientific analysis, including the
transit detection, is performed. These data will typically be
available a few months after the end of each run.  However, a
preliminary real-time analysis of the N1 light curves is carried out
to detect interesting events (for example, events which could be
transits of giant planets). The purpose of this `alarm mode' is to
provide a weekly update of the list of targets to be oversampled
(i.e.\ targets for which data is downlinked to Earth every 32 rather
than 512\,s.)

\subsection{Blind tests}

The detection of transits in the CoRoT light curves, and particularly
of shallow transits caused by terrestrial planets, is a challenging
task. First one must minimise the impact of a number of high- and
low-frequency noise sources affecting the light curves, which become
important at the level of precision achieved with CoRoT. Data
taken during passages through the South Atlantic Anomaly (SAA)
unusable, and the light curves therefore contain a few gaps lasting
approximately 10 min during each 24\,h period. The data around these
gaps are also affected by higher background noise level. Additionally,
all stars are expected to display some degree of variability
(micro-variability, associated with the rotational modulation and
intrinsic evolution of surface features of magnetic or convective
origin), which must be treated before the shallowest and most
interesting transits can be detected. 

In order to prepare for the analysis of the CoRoT light curves, the
CoRoT Exoplanet Science Team (CEST) has conducted two blind tests. In
both cases, a `game master' generated a set of light curves including
as many of the foreseen noise sources as possible, using a
detailed model of the CoRoT instrument and the micro-variability
models of \citet{aig_lrp04} and \citet{aig_afg04}. Simulated transit
events and a range of astrophysical contaminants (events which mimic
transits but are not of planetary origin) were then added some of the
light curves before distributing them to the participating analysis
teams, but the true content of the light curves was known only to the
game master.

The first blind test (BT1, \citealt{aig_mpb+05}) was focussed on
detection, and so a relatively large sample of light curves was
generated with few containing transit-like events. Only white light
light curves were simulated for this first exercise.  Five teams took
part (two from the Laboratoire d'Astrophysique de Marseille, one from
Geneva Observatory, one from the Institute for Planetary research in
Berlin and one from the Institute of Astronomy in Cambridge), and the
main results of the exercise were the following:
\begin{itemize}
\item Micro-variability can be filtered out relatively effectively
  without strongly affecting the transits by exploiting the distinct
  frequency signatures of the two types of signal: micro-variability
  is strongest on timescales of days and longer, while transits occur
  over timescales of minutes (ingress/egress) to hours.
\item A number of detection algorithms were tested, including matched
  filters, and cross-correlation methods, all of which were relatively
  successful, but the best performances in terms of sensitivity and
  false alarm rejection were obtained with algorithms based on
  least-squares fitting of box-shaped transits, such as the well-known
  BLS \citep{aig_kzm02}.
\item No false alarms were reported by more than one team,
  highlighting the advantage of several teams working with different
  methods but sharing their results.
\item By tracing a line on a diagram of transit depth $d$ versus
  number $n$ of observed transits which separates events
  detected by most teams from events missed by most teams, a rough
  detection limit of $d \sim 2 \times 10^{-3} n^{-1/2}$ was
  established. This dependence on $n$ is as expected if the number of
  observations per transit and the noise level per observation is
  roughly the same for all cases, although in reality a number of
  other factors, in particular star magnitude and activity level, also
  affect the detection limit.
\end{itemize}

\noindent The second blind test (BT2, \citealt{aig_ma07}) was focussed
mainly on discriminating between real transit events and transit-like
events of stellar origin, often referred to as astrophysical false
positives, and therefore a smaller number of light curves were generated
but all contained a signal -- transit or false positive. The transit or
eclipse light curve alone does not enable the companion's mass to be
measured, and spectroscopic follow-up is always necessary to confirm
the planetary nature of any candidate, but a significant fraction of
the astrophysical false positives can be eliminated by a close inspection
of the light curve, including the colour information.

An important aspect of BT2 was testing the use of the information
which will be known about each target in advance as a result of the
preliminary ground-based observations (see
Section~\ref{aig_strategy}). These observations provide a list of
background stars which contribute some flux to the aperture of each
target in the CoRoT fields, or `contaminants'. This information is
stored, along with the results of other preliminary observations and,
in the long run, the results of follow-up observations aimed at
obtaining precise parameters for the host stars of the best planet
candidates, in a dedicated database called EXODAT \citep{aig_dmd+06},
which will be accessible to the scientist analysing the CoRoT light
curves.

The exercise is still ongoing, but preliminary conclusions include:
\begin{itemize}
\item Direct application of simple diagnostics to distinguish between
  planetary transits and stellar eclipses based on event duration
  (such as the $\eta$ parameter proposed by \citealt{aig_ts05}) or colour
  (assuming planetary transits are grey, whereas most stellar eclipses
  are not) can lead to the rejection of real transit events, and
  should be used with care.
\item Diagnostics using the presence of secondary eclipses or
  ellipsoidal variability are robust and can be used to distinguish
  some stellar binaries from planets.
\item Knowledge of the positions and colours of the `contaminants' can
  be used along with the light curve (white light alone or including
  colour information) to identify many likely blended eclipsing binaries
  (stellar binaries whose signal is diluted by a third, constant light
  source and therefore resembles a planetary transit).
\item The most difficult type of false positive to identify from the
  light curves and contaminants information alone is the case of
  stellar eclipses with a small (M-type) secondary, but this type of
  system has a clear radial velocity signature easily identified from
  a few spectroscopic observations.
\end{itemize}

\subsection{Estimating the planet yield and follow-up needs}

The samples of planets and eclipsing binaries included in the blind
tests were designed to contain examples of all the types of events
expected, but were not intended to reproduce the real relative
frequencies of the different events, or the incidence of transit-like
event in the entire sample of light curves. Thus , they tell us what
kind of events are detectable and characterisable, but not how many of
them to expect.

CoRoTLux \citep{aig_fgm+07} is a tool to evaluate the yield of transit
surveys which simulates the entire stellar population observed
(including background stars) based on standard Galactic models
\citep{aig_rrd+01} and generates a population of eclipsing binaries and of
planets using the most up-to date binarity and planet incidence
statistics. It simulates light curves with realistic astrophysical
and instrumental noise sources, after which a matched filter transit
detection algorithm is used to evaluate the number of candidate
transits expected. It can be applied to a number of ground or
space-based surveys. 

In the case of CoRoT, the Galactic model used is adjusted to match the
star counts in EXODAT. Two sets of CoRoTLux simulations were carried
out: one to investigate the yield of giant planets, for which the
incidence and orbital period distribtions are relatively well known,
and one to investigated the yield of `terrestrial' planets ($\sim
3$--$4\,R_{\oplus}$), for which the incidence is unknown, but was
assumed to be 5 times higher than that of the giants\footnote{The
  motivation behind this assumption is that less solid material is
  needed to form a terrestrial planet, and that rapid growth is less
  crucial than for a giant, which much reach the critical mass for
  runaway accretion before the gas in the proto-planetary disk
  disappears.}. The results will be described in detail in a
forthcoming paper (Fressin et al.\ in prep.), but preliminary
indications are that CoRoT should detect $\sim 7$ giant planets per
run and $\sim3$--4 terrestrial planets per long run, i.e.\ a total of
$\sim 65$ giant planets and $\sim 20$ terrestrial
planets. Approximately 100 transit candidates per long run are
foreseen, of which the BT2 results to date suggest 50\% may be
identifiable as stellar events from the light curves and EXODAT
information alone. The shallowest transits, which are potentially the
most interesting, carry the highest follow-up cost, as the relative
fraction of true transits to stellar mimics decreases strongly with
decreasing transit depth.

\section{Follow-up}

\begin{figure}[t]
  \centering
  \plotfiddle{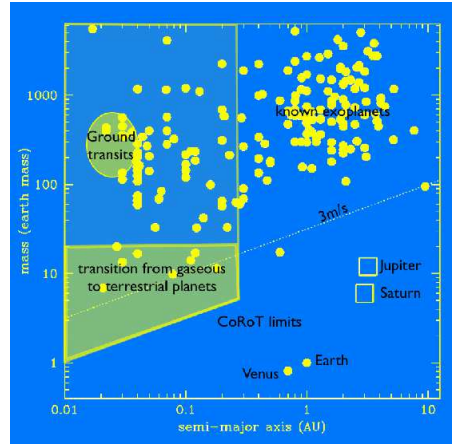}{8cm}{0.0}{30}{30}{-120}{0}
  \caption{Detection capabilities of CoRoT compared to known exoplanets (dots) and present day radial velocity facilities (the dashed diagonal line represents the 3\,m/s limit, the best instruments in the world currently achieve 1\,m/s consistently). The shaded ellipse shows the approximate locus of transiting planets detected from the ground, while the shaded trapeze shows the transition regime between gaseous giant and terrestrial planets which CoRoT will be the first to explore. Solar system planets are shown for comparison in the bottom right part of the diagram.
  \label{aig_follow}}
\end{figure}

Three main types of follow-up observations are foreseen:
\begin{itemize}
\item Photometric follow-up, to obtain high spatial resolution images
  in- and out-of transit in order to identify which of the stars
  falling in the CoRoT aperture is being eclipsed, and thus what is
  the true (undiluted) depth of the event. As the PSF of CoRoT is
  large, 1\,m class telescopes should suffice to obtain a significant
  resolution improvement. Among the facilities foreseen for this
  follow-up are the 80\,cm telescope at Observatorio del Teide in
  Tenerife, and the Euler telescope in La Silla.
\item Radial velocity follow-up, to measure the companion mass. To
  make optimal use of the available facilities, a progressive strategy
  is foreseen, involving initial follow-up wit FLAMES in Paranal for
  the fainter objects and, for the brighter candidates, with CORALIE
  in la Silla, SOPHIE at the Observatoire de Haute Provence, and the
  2\,m telescope at the Th\"uringer Landessternwarte Tautenburg. The
  best low radial velocity amplitude candidates will then be targeted
  with HARPS in la Silla. As illustrated on Figure~\ref{aig_follow},
  CoRoT's detection capabilities are particularly well-matched to the
  world-leading radial velocities currently available;
\item High-resolution spectra of the host stars, using e.g.\ UVES on
  the VLT, to derive accurate physical parameters (for confirmed 
  candidates only).
\end{itemize}
To ensure prompt confirmation of transit candidates, the activities of
the four CEST teams in charge of light curve analysis, photometric
follow-up, radial velocity follow-up and stellar parameter
determination will be co-ordinated by a small team which will
regularly re-analyse the available data for each candidate in real
time and review the candidate priority rankings for each type of
follow-up.  Once candidates are confirmed, further follow-up of the
giant planets is foreseen using space based facilities, with the goal
of obtaining extended baseline and improved precision light curves
(HST: improved radii, search for other non-transiting planets by
transit timing, search for rings and moons, oblateness measurements)
and detecting secondary eclipses in the infrared (Spitzer: surface
temperature measurement and atmospheric studies).

\section{The additional program}
\label{aig_ap}

Although primarily driven by asteroseismology and exoplanet detection,
CoRoT also accomodates an additional program, that
includes all scientific use of CoRoT data for purposes outside the
core program areas of exoplanet detection in the exoplanet field and
asteroseismology in the asteroseismology field. Thus, exoplanet
detection in the asteroseismology field and asteroseismology in the
exoplanet field, as well as any other analysis of CoRoT light curves
(rotation, activity, flaring, eclipsing binaries) fall under the
additional program.

Through yearly Announcements of Opportunity (AOs), members of the
astronomical community at large in the participating countries
(including all ESA member states) can propose either to systematically
analyse core program data for scientific purposes outside the core
program, or to observe specific targets in the CoRoT
fields. Additionally, a small number of short runs will be driven
primarily by the additional program. One such run is foreseen on the
NGC\,2264 star forming region. The AO for the first year of CoRoT
operations is past, but an announcement for the second year is
foreseen early in 2007.

\section{Conclusions}
\label{aig_concl}

CoRoT is the first space mission dedicated to exoplanet detection, and
the first project capable of detecting transits of terrestrial
planets. While stopping short of detecting habitable planets, its
capabilities will enable it to explore for the first time the
uncharted transition regime between the giant and terrestrial
planets found in our own system.

Prompt follow-up is expected to enable the confirmation of the first
candidate planets from the initial run and the first long run (alarm
mode) by the summer of 2007. As shallow transits will be detectable
only in the full long run N2 data, the first of which are expected
around August 2007, the first confirmed terrestrial planets are
expected in the spring of 2008. The CoRoT light curves will be come
public one year after their release to the Co-Is, i.e.\ approximately
15 months after the end of each run.

For more information, the reader is referred to the following sources:
\begin{itemize}
\item the official CoRoT website: {\tt http://corot.oamp.fr};
\item \emph{The CoRoT mission: pre-launch status -- Stellar seismology
    and planet finding}: ESA-SP 1306, eds.\ M.\ Fridlund, A.\ Baglin,
  L.\ Conroy \& J.\ Lochard (2006)\footnote{Available on request from
    M.\ Fridlund, {\tt malcolm.fridlund@esa.int}.};
\item for information on the additional program: \\
{\tt http://ams.astro.univie.at/?s=space;corotAPWG}.
\end{itemize}

\acknowledgements The contents of this paper are the result of
contributions from all members of the CoRoT community, and in
particular from the CoRoT Exoplanet Science Team.

\end{document}